\documentclass[twocolumn]{openjournal}
\usepackage{xcolor}
\usepackage{textgreek}
\usepackage[utf8]{inputenc}
\usepackage[english]{babel}
\usepackage{hyperref}
\hypersetup{
    unicode, 
    colorlinks=true,
    linkcolor=linkcolor,
    citecolor=linkcolor,
    filecolor=linkcolor,
    urlcolor=linkcolor,
}
\usepackage{color,colortbl}
\definecolor{linkcolor}{rgb}{0.0,0.3,0.5}
\usepackage{tensind}
\tensordelimiter{?}
\DeclareGraphicsExtensions{.bmp,.png,.jpg,.pdf}
\usepackage{verbatim}
\usepackage[normalem]{ulem}
\usepackage{orcidlink}
\usepackage{soul}
\usepackage{amsmath}
\usepackage{savesym}
\savesymbol{tablenum}
\usepackage{siunitx}
\DeclareSIUnit\solarmass{M_{\odot}}
\restoresymbol{SIX}{tablenum}

\newcommand{\primary}{\ensuremath{1}}
\newcommand{\secondary}{\ensuremath{2}}
\makeatletter
\providecommand*{\diff}{\@ifnextchar^{\DIfF}{\DIfF^{}}}
\def\DIfF^#1{%
    \mathop{\mathrm{\mathstrut d}}\nolimits^{#1}\gobblespace
}
\def\gobblespace{%
    \futurelet\diffarg\opspace
}
\def\opspace{%
    \let\DiffSpace\!%
    \ifx\diffarg(%
        \let\DiffSpace\relax
    \else
        \ifx\diffarg[%
            \let\DiffSpace\relax
        \else
            \ifx\diffarg\{%
                \let\DiffSpace\relax
            \fi\fi\fi\DiffSpace
}

\begin{document}

\title{A method for constructing the joint mass function of binary stars}

\author{Amery Gration}
\email{a.gration@surrey.ac.uk}
\affiliation{School of Mathematics and Physics, University of Surrey, Guildford, GU2 7XH, UK}

\author{Robert G. Izzard}
\affiliation{School of Mathematics and Physics, University of Surrey, Guildford, GU2 7XH, UK}

\author{Payel Das}
\affiliation{School of Mathematics and Physics, University of Surrey, Guildford, GU2 7XH, UK}

\begin{abstract}
\noindent
The initial mass function (IMF) describes the distribution of stellar masses in a population of newly born stars and is amongst the most fundamental concepts in astrophysics. It is not only the direct result of the star formation process but it also explains the evolution of galaxies' luminosities, metal yields, star-formation efficiencies, and supernova production rates. Because most stars exist in binary systems, however, a full statistical account of stellar mass requires not the IMF but rather the joint distribution of a binary population's primary- and secondary-star masses. This joint distribution must respect the IMF of the stars from which the population has been assembled as well as the distribution of mass ratios that results from the assembly mechanism. Despite its importance, this joint distribution is known only in the case of random pairing. Here we present a method for constructing it in the general case. We also illustrate the use of our method by recovering the known result for random pairing and by finding the previously unknown result for uniform pairing.
\end{abstract}

\maketitle

\section{Introduction}
\label{sec:org003c235}

The mass of a star determines its physical properties, such as radius and luminosity, and consequently its subsequent evolution including its final fate as a white dwarf, neutron star, or black hole.
In turn, the distribution of stellar masses in a given population of stars determines the current and future properties of that population \citep{miller1979,scalo1986,kroupa2002,chabrier2003}.
For example, the distribution of stellar masses in a galaxy determines its luminosity as well as its future metal yields, star-formation efficiency, and supernova production rate.

The distribution of stellar masses in a population is described by the mass function (MF), which is the probability density function (PDF) of stellar mass treated as a random variable.
The initial mass function (IMF) describes the distribution of stellar masses in a population of newly born stars whilst the present-day mass function (PDMF) describes the distribution of masses in some extant stellar population.
In general such a population has a complex star formation history and therefore consists of multiple subpopulations of coeval stars.
But even for a subpopulation of this kind the PDMF differs from the IMF due to stellar evolution. In providing the initial conditions for this evolution the IMF is crucial in explaining a stellar population's properties.

Most stars exist in binary systems \citep{offner2023} and a full statistical account of their masses therefore requires the joint distribution of primary- and secondary-star masses.
This joint distribution determines the distribution of the masses of the components of a population of binary systems as well as the conditional distribution of their mass ratios given their primary-star mass, which is itself amenable to observation \citep{offner2023}.
Knowledge of the joint distribution of primary- and secondary-star masses is crucial in correctly inferring the IMF from observations and in modelling the evolution of binary-star populations using population synthesis \citep{kroupa1991,izzard2019}.
But despite its importance there is no method for constructing a joint distribution that respects a particular distribution of stellar masses and a particular conditional distribution of mass ratios given primary-star mass. This joint distribution is known only in the case of random pairing, where two stars are drawn randomly from a population of single stars \citep{malkov2001}. In all other cases approximate alternatives are unavoidable \citep{kouwenhoven2009}.

Here we present a method for constructing the joint distribution of primary- and secondary-star masses in the general case. To determine this distribution it is in fact sufficient to determine the PDF of primary-star mass. We therefore write down an integral equation in which this PDF appears as the unknown function. We then illustrate the use of our method by solving the equation in two special cases. First, we recover the known result for random pairing. Secondly, we find the unknown result for uniform pairing, in which the conditional distribution of the mass ratio given the primary mass is uniform on some interval.

\section{Constructing the joint-mass function}
\label{sec:org01b6739}

Consider a population of binary star systems and suppose that we could ionize the members of this population to produce a population of free stars.
We may treat the masses of these free stars as independent and identically distributed (IID) continuous random variables. A star sampled from this population has a mass \(M\) with range \((m_{\min}, m_{\max})\) and PDF \(f_{M}\), which is the MF.\footnote{We will follow the mathematicians' convention of using an upper case letter to denote a random variable but a lower case letter to denote the realization of a random variable. Here, the random variable \(M\) may have a realization \(m\). It has PDF \(f_{M}\), which has value \(f_{M}(m)\) at \(m\).} The masses of primary stars in binary systems, considered as a distinct population, are also IID continuous random variables, as are the masses of secondary stars, again considered as a distinct population. A star sampled from the population of primary stars has a mass \(M_{\primary}\) with PDF \(f_{M_{\primary}}\), which we will call the `primary mass function' (PMF). Similarly, a star sampled from the population of secondary stars has a mass \(M_{\secondary}\) with PDF \(f_{M_{\secondary}}\), which we will call the `secondary mass function' (SMF).

We may treat the pairs of stellar masses in the population of binary systems as IID random vectors. A binary system sampled from this population has a pair of masses \((M_{\primary}, M_{\secondary})\). The elements of this pair have a joint probability density function \(f_{(M_{\primary}, M_{\secondary})}\), which we will call the `joint mass function' (JMF). The mass ratio of the binary binary system is the random variable \(Q := M_{\secondary}/M_{\primary}\) and the conditional PDF of \(Q\) given \(M_{\primary}\) is \(f_{Q|M_{\primary}}\), which we will call the `conditional mass-ratio function' (CMRF). Knowing the CMRF is equivalent to knowing the conditional PDF of \(M_{\secondary}\) given \(M_{\primary}\), \(f_{M_{\secondary}|M_{\primary}}\), which we will call the `conditional secondary mass function' (CSMF) since, by a change of variables, 
\begin{align}
\label{eq:pairing_function}
f_{M_{\secondary}|M_{\primary}}(m_{\secondary}|m_{\primary}) = \dfrac{1}{m_{\primary}}f_{Q|M_{\primary}}(m_{\secondary}/m_{\primary}|m_{\primary})
\end{align}
or, equivalently,
\begin{align}
f_{Q|M_{\primary}}(q|m_{\primary}) = m_{\primary}f_{M_{\secondary}|M_{\primary}}(qm_{\primary}|m_{\primary}).
\end{align}

\subsection{Constructing the primary-mass function}
\label{sec:orgf22604b}

Our task is to find a JMF that is consistent with a given MF and a given CSMF. To do so we note that it suffices to find the PMF, \(f_{M_{\primary}}\), since
\begin{align}
\label{eq:conditional_pdf}
f_{(M_{\primary}, M_{\secondary})}(m_{\primary}, m_{\secondary}) = f_{M_{\secondary}|M_{\primary}}(m_{\secondary}|m_{\primary})f_{M_{\primary}}(m_{\primary}).
\end{align}
To find the PMF, then, note that (i) the distribution of \(M\) is a mixture of the distributions of \(M_{\primary}\) and \(M_{\secondary}\) and (ii) \(M_{\secondary}\) may be considered to be the marginal variable of the pair \((M_{\primary}, M_{\secondary})\). Then
\begin{align}
\label{eq:mixture_model}
f_{M}(m) = \dfrac{1}{2}(f_{M_{\primary}}(m) + f_{M_{\secondary}}(m))
\end{align}
and
\begin{align}
\label{eq:secondary_mass_pdf}
f_{M_{\secondary}}(m_{\secondary}) =
\int_{m_{\min}}^{m_{\max}}
f_{(M_{\primary}, M_{\secondary})}(m_{\primary}, m_{\secondary})\diff{}m_{\primary}.
\end{align}
Using Equation~\ref{eq:conditional_pdf} we find that
\begin{align}
\label{eq:fredholm_equation_for_primary_mass_function}
f_{M_{\primary}}(m) = 2f_{M}(m) - \int_{m_{\min}}^{m_{\max}}f_{M_{\secondary}|M_{\primary}}(m|m_{\primary})f_{M_{\primary}}(m_{\primary})\diff{}m_{\primary}.
\end{align}
This is an inhomogeneous Fredholm integral equation of the second kind with kernel \(f_{M_{\secondary}|M_{\primary}}\) and known function \(f_{M}\). Its solution is the PMF, \(f_{M_{\primary}}\). By finding this solution we find the JMF, again by Equation~\ref{eq:conditional_pdf}.

The theory of Fredholm integral equations and their solution is well established \citep{porter1990,hackbusch1995,atkinson1997}. Depending on the formula for the CSMF, we may solve Equation~\ref{eq:fredholm_equation_for_primary_mass_function} in closed form, in the form of an infinite series, or by using numerical methods. Note, however, that we have constructed Equation~\ref{eq:fredholm_equation_for_primary_mass_function} on the assumption that the MF, CSMF, and PMF are not only PDFs but that there exists a JMF from which the MF, CSMF, and PMF may be derived. In practice we do not know the MF and CSMF but instead have more-or-less well-motivated formulas for them. In these circumstances such a JMF may not exist. Equation~\ref{eq:fredholm_equation_for_primary_mass_function} may not admit a solution or may admit a solution that is not a PDF, which must be non-negative and integrate to one.

The non-negativity is the problem. Any solution necessarily integrates to one, which we can see by integrating both sides of Equation~\ref{eq:fredholm_equation_for_primary_mass_function} with respect to \(m\) and using the fact that
\begin{align}
\int_{m_{\min}}^{m_{\max}}f_{M_{\secondary}|M_{\primary}}(m|m_{\primary})f_{M_{\primary}}(m_{\primary})\diff{}m = 1
\end{align}
for all \(m_{\primary}\) whereupon
\begin{align}
\int_{m_{\min}}^{m_{\max}}f_{M_{\primary}}(m)\diff{}m
=
2 - \int_{m_{\min}}^{m_{\max}}f_{M_{\primary}}(m_{\primary})\diff{}m_{\primary},
\end{align}
and hence
\begin{align}
\int_{m_{\min}}^{m_{\max}}f_{M_{\primary}}(m)\diff{}m
=
1.
\end{align}
We must, therefore, always check that the solution is non-negative. If a solution is not non-negative then the chosen MF and CSMF are inconsistent and a JMF that respects them both does not exist.

\subsection{Main-sequence binary systems}
\label{sec:orgce03346}

In general the mass of the secondary star can be greater than the mass of the primary star. However, in binary systems consisting of two main-sequence stars the mass of the secondary star is in fact always less that the mass of the primary star. In these circumstances we can investigate the limiting behaviour of the PMF and SMF without solving Equation~\ref{eq:fredholm_equation_for_primary_mass_function}. For main-sequence binary systems the PMF is given by 
\begin{align}
f_{M_{\primary}}(m_{\primary})
=
\int_{m_{\min}}^{m_{\primary}}
f_{(M_{\primary}, M_{\secondary})}(m_{\primary}, m_{\secondary})\diff{}m_{\secondary}
\end{align}
and therefore vanishes in the low-mass limit, i.e.
\begin{align}
\lim_{m_{\primary} \longrightarrow m_{\min}}f_{M_{\primary}}(m_{\primary}) = 0.
\end{align}
Consequently (by Equation~\ref{eq:mixture_model}) the value of the SMF is twice the value of the MF, i.e.
\begin{align}
\lim_{m_{\secondary} \longrightarrow m_{\min}^{+}}f_{M_{\secondary}}(m_{\secondary}) = 2f_{M}(m_{\min}).
\end{align}
(Note that we consider the limit of \(f_{M_{\secondary}}\) as \(m_{\secondary}\) approaches \(m_{\min}\) from above since \(f_{M_{\secondary}}\) is discontinuous at \(m_{\min}\).) Similarly, in main-sequence binary systems the SMF is given by 
\begin{align}
f_{M_{\secondary}}(m_{\secondary})
&=
\int_{m_{\secondary}}^{m_{\max}}
f_{(M_{\primary}, M_{\secondary})}(m_{\primary}, m_{\secondary})\diff{}m_{\primary}
\end{align}
and therefore vanishes in the high-mass limit, i.e.
\begin{align}
\lim_{m_{\secondary} \longrightarrow m_{\max}}f_{M_{\secondary}}(m_{\secondary}) = 0.
\end{align}
Consequently (by Equation~\ref{eq:mixture_model}) the value of the PMF is twice the value of the MF, i.e.
\begin{align}
\lim_{m_{\primary} \longrightarrow m_{\max}^{-}}f_{M_{\primary}}(m_{\primary}) = 2f_{M}(m_{\max}).
\end{align}
(Note that we consider the limit of \(f_{M_{\primary}}\) as \(m_{\primary}\) approaches \(m_{\max}\) from below since \(f_{M_{\primary}}\) is discontinuous at \(m_{\max}\).) In other words, there are no secondary stars less massive than the least-massive primary star and there are no primary stars more massive than the most-massive secondary star.

\section{Random and uniform pairing}
\label{sec:org6db38cd}

With Equation~\ref{eq:fredholm_equation_for_primary_mass_function} in hand we will now construct the JMF in the two cases of random and uniform pairing. In random pairing, the components of a binary system are drawn from a pool of free stars and the more-massive star designated the primary. In uniform pairing, the components are drawn so that the conditional distribution of \(Q\) given \(M_{\primary}\) is uniform on some interval. That interval itself depends on \(M_{\primary}\). To set its infimum we will require that \(Q\) is never less than \(q_{\min}\).\footnote{The infimum and supremum of an interval are its greatest lower bound and its least upper bound.} The existence of a minimum stellar mass places an additional constraint on \(Q\), namely that it is also never less than \(m_{\min}/m_{\primary}\). The infimum of the interval is therefore \(\max(q_{\min}, m_{\min}/m_{\primary})\). We will also require that the secondary mass never exceeds the primary mass and hence that the supremum is 1.

In both cases of random and uniform pairing we will write the CSMF in its general form and solve Equation~\ref{eq:fredholm_equation_for_primary_mass_function} to find the PMF and hence the JMF. The PMF and JMF for random pairing are already known in closed form \citep{malkov2001}. Our aim here is to show that our method recovers the known results before we proceed to the novel case of uniform pairing.

By way of providing concrete examples we will assume that the MF is equal to the canonical IMF (i.e.~the IMF of the Solar neighbourhood), \(\xi\), and that this is a split-power law given by
\begin{align}
\label{eq:kroupa_imf}
\xi(m)
=
A
\begin{cases}
m^{-1.3}   &\text{ if $m \in [m_{\min}, 0.5)$,}\\
0.5m^{-2.3} &\text{ if $m \in [0.5, m_{\max}]$}
\end{cases}
\end{align}
where
\begin{align}
\dfrac{1}{A}
=
\dfrac{0.5(0.5^{-1.3} - m_{\max}^{-1.3})}{1.3}
+ \dfrac{m_{\min}^{-0.3} - 0.5^{-0.3}}{0.3}
\end{align}
and where \(m\) is in units of solar mass \citep{kroupa2001}. Furthermore, we will assume that \(m_{\min} = \qty{0.08}{\solarmass}\) (the hydrogen-burning limit and hence the minimum possible stellar mass) and \(m_{\max} = \qty{150}{\solarmass}\) (a plausible maximum stellar mass). 

\subsection{Random pairing}
\label{sec:orgcc3815a}

To find the CSMF for random pairing we work backwards from the already-known JMF, which is given by 
\begin{align}
\label{eq:joint_mass_function_random}
f_{(M_{\primary}, M_{\secondary})}(m_{\primary}, m_{\secondary}) = 2f_{M}(m_{\primary})f_{M}(m_{\secondary})
\end{align}
with support\footnote{The support of a PDF is the interval on which it is nonzero.} on the closed lower triangle
\begin{align}
((m_{\primary}, m_{\secondary}): (m_{\primary}, m_{\secondary}) \in I^{2} \wedge m_{\secondary} \le m_{\primary})
\end{align}
where \(I := (m_{\min}, m_{\max})\) \citep{malkov2001}.
This immediately yields the PMF, which is given by
\begin{align}
f_{M_{\primary}}(m_{\primary})
&= \int_{m_{\min}}^{m_{\max}}f_{(M_{\primary}, M_{\secondary})}(m_{\primary}, m_{\secondary})\diff{}m_{\secondary}\\
\label{eq:pmf_integral}
&= 2f_{M}(m_{\primary})\int_{m_{\min}}^{m_{\primary}}f_{M}(m_{\secondary})\diff{}m_{\secondary}\\
\label{eq:primary_mass_pdf_random}
&=
2f_{M}(m_{\primary})F_{M}(m_{\primary}),
\end{align}
with support on the interval \(I\) and where \(F_{M}\) is the cumulative distribution function of \(M\). But to illustrate our method we require the CSMF which, by definition of the conditional PDF (Equation~\ref{eq:conditional_pdf}), is given by
\begin{align}
\label{eq:pairing_function_random}
f_{M_{\secondary}|M_{\primary}}(m_{\secondary}|m_{\primary})
&=
\dfrac{f_{(M_{\primary}, M_{\secondary})}(m_{\primary}, m_{\secondary})}{f_{M_{\primary}}(m_{\primary})}\\
&=
\dfrac{2f_{M}(m_{\primary})f_{M}(m_{\secondary})}{2f_{M}(m_{\primary})F_{M}(m_{\primary})}\\
\label{eq:random_csmf}
&=
\dfrac{f_{M}(m_{\secondary})}{F_{M}(m_{\primary})},
\end{align}
with support on the closed upper triangle
\begin{align}
\label{eq:support_of_random_csmf}
((m_{\secondary}, m_{\primary}): (m_{\secondary}, m_{\primary}) \in I^{2} \wedge m_{\secondary} \le m_{\primary}).
\end{align}

We see, by substitution, that Equation~\ref{eq:primary_mass_pdf_random} then solves our integral equation (Equation~\ref{eq:fredholm_equation_for_primary_mass_function}) although we may also solve it directly. The SMF follows by Equation~\ref{eq:mixture_model} and is given by
\begin{align}
\label{eq:secondary_mass_pdf_random}
f_{M_{\secondary}}(m_{\secondary})
= 2f_{M}(m_{\secondary})(1 - F_{M}(m_{\secondary})),
\end{align}
again with support on the interval \(I\). Note that the PMF has a maximum when
\begin{align}
f_{M}(m)^{2} = -F_{M}(m)\dfrac{\diff{}f_{M}(m)}{\diff{}m},
\end{align}
which only occurs where the derivative of the MF is negative. The PMF, SMF, and MF are equal when \(F_{M}(m) = 1/2\). For masses greater than this \(f_{M_{\primary}}(m) > f_{M}(m) > f_{M_{\secondary}}(m)\) whilst for masses less than this \(f_{M_{\primary}}(m) < f_{M}(m) < f_{M_{\secondary}}(m)\).

\begin{figure}{}
\centering
\includegraphics[width=8.cm]{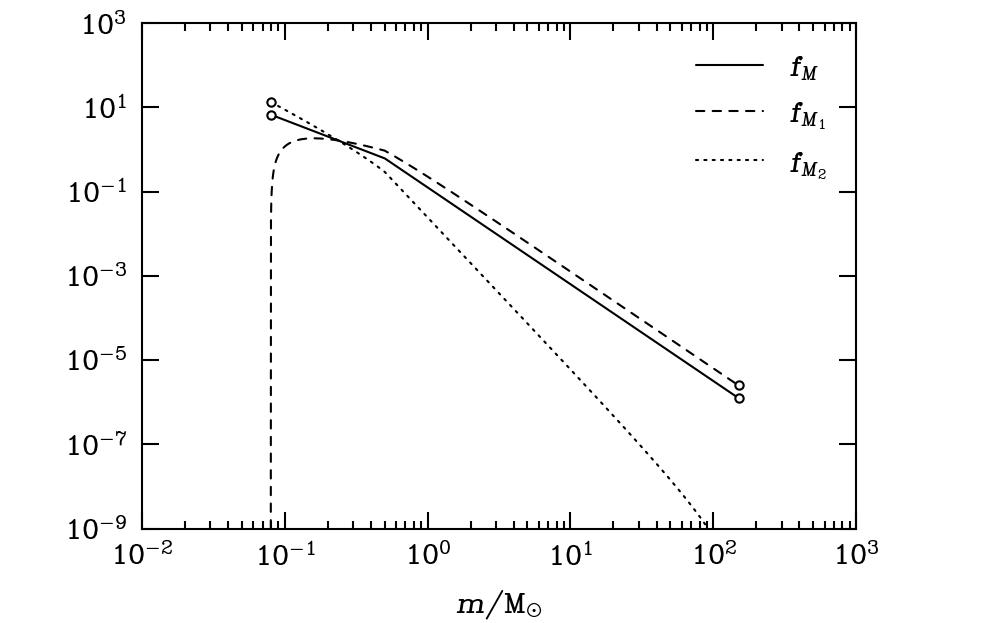}
\caption{\label{fig:org91e602c}The primary-mass function, \(f_{M_{\primary}}\) (Equation~\ref{eq:primary_mass_pdf_random}), and secondary-mass function, \(f_{M_{\secondary}}\) (Equation~\ref{eq:secondary_mass_pdf_random}), alongside the mass function, \(f_{M}\) (Equation~\ref{eq:kroupa_imf}). Here we have used the IMF of \citet{kroupa2001} and random pairing (Section~\ref{sec:orgcc3815a}).}
\end{figure}

\begin{figure}{}
\centering
\includegraphics[width=8.cm]{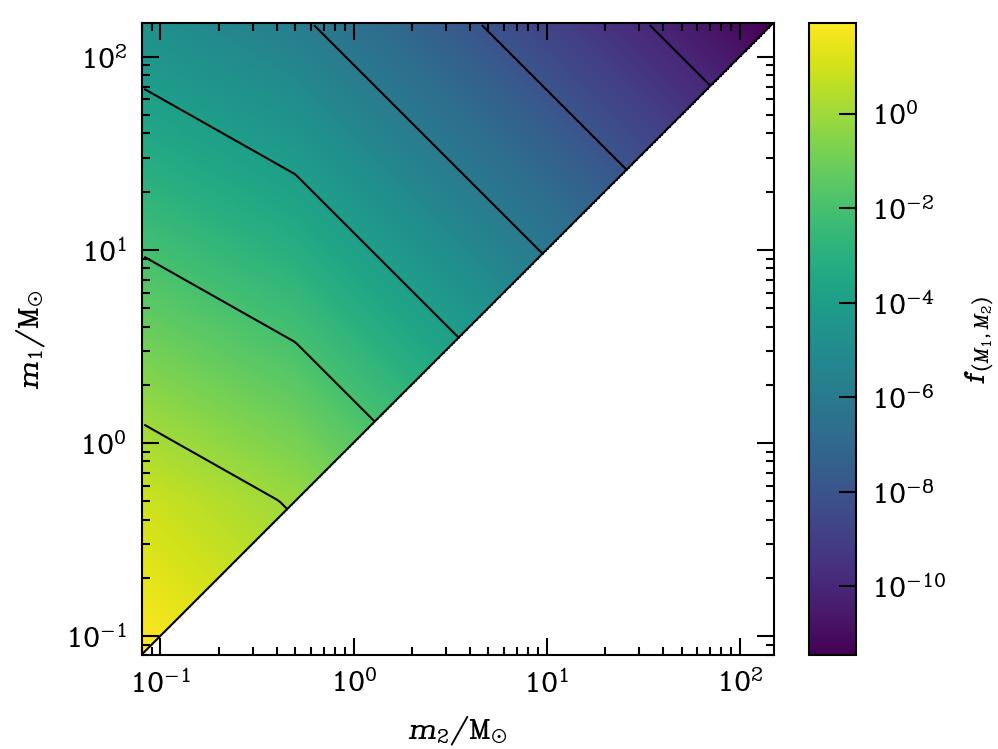}
\caption{\label{fig:org0ab2b78}The joint mass function, \(f_{(M_{\primary}, M_{\secondary})}\) (Equation~\ref{eq:joint_mass_function_random}), constructed using the IMF of \citet{kroupa2001} and random pairing (Section~\ref{sec:orgcc3815a}, Figure~\ref{fig:org91e602c}). It is non-zero for \(m_{\secondary} \le m_{\primary}\). Low-mass pairs are most probable and high-mass pairs least probable.}
\end{figure}

Now let us suppose that \(f_{M} = \xi\). We plot the PMF and SMF alongside the MF in Figure~\ref{fig:org91e602c} and the JMF in Figure~\ref{fig:org0ab2b78}. The PMF has a maximum at \(m = 0.08(1.3/1.6)^{-1/0.3}\unit{\solarmass} = \qty{0.16}{\solarmass}\) while the SMF and MF have maxima at \(m = \qty{0.08}{\solarmass}\). The three functions are equal at \(m = (0.08^{-1/0.3} - 0.15/A)\unit{\solarmass} = \qty{0.24}{\solarmass}\).

\subsection{Uniform pairing}
\label{sec:org96ac7dd}

To find the CSMF for uniform pairing we begin with the CMRF, which is given by
\begin{align}
\label{eq:uniform_cmrf}
f_{Q|M_{\primary}}(q|m_{\primary}) = \dfrac{1}{1 - \max(q_{\min}, m_{\min}/m_{\primary})}
\end{align}
with support on the closed region
\begin{align}
((q, m_{\primary}): (q, m_{\primary}) \in J \times I \wedge qm_{\primary} \ge m_{\min})
\end{align}
where \(J := (q_{\min}, 1]\). By Equation~\ref{eq:pairing_function} we then find that the CSMF is given by
\begin{align}
\label{eq:uniform_csmf}
f_{M_{\secondary}|M_{\primary}}(m_{\secondary}|m_{\primary}) = \dfrac{1}{m_{\primary}(1 - \max(q_{\min}, m_{\min}/m_{\primary}))}
\end{align}
with support on the closed diagonal band
\begin{align}
\label{eq:support_of_uniform_csmf}
((m_{\secondary}, m_{\primary}): (m_{\secondary}, m_{\primary}) \in I^{2} \wedge m_{\secondary} \le m_{\primary} \wedge m_{\secondary} \ge q_{\min}m_{\primary}).
\end{align}
Given this CSMF we cannot find the PMF in closed form. Instead, we must compute it numerically (see \hyperref[sec:org5b48980]{Appendix}).

\begin{figure}{}
\centering
\includegraphics[width=8.cm]{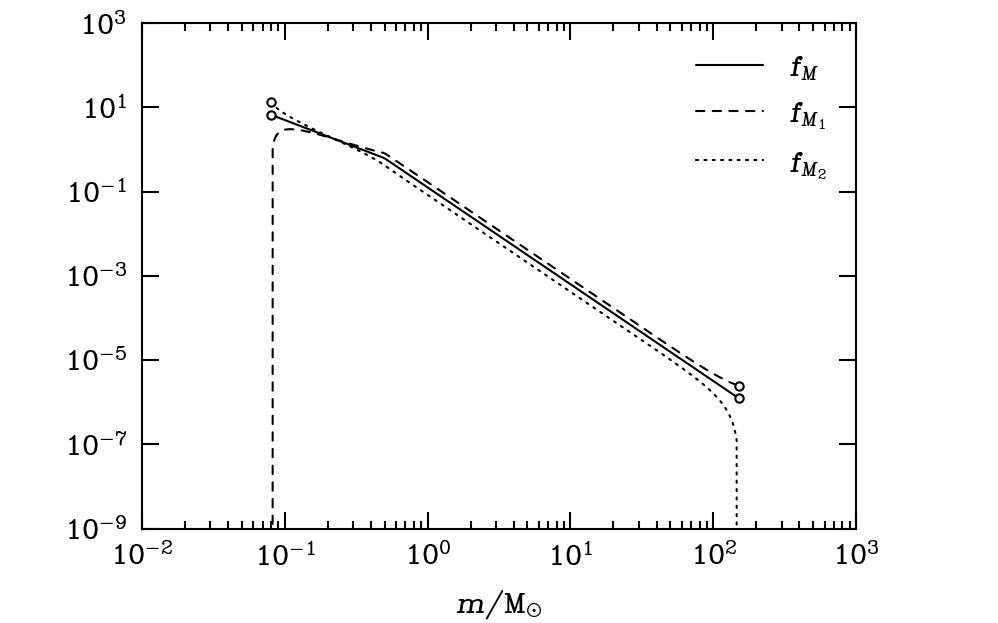}
\caption{\label{fig:org1e8a02d}The primary-mass function, \(f_{M_{\primary}}\), and secondary-mass function, \(f_{M_{\secondary}}\), alongside the mass function, \(f_{M}\) (Equation~\ref{eq:kroupa_imf}). Here we have used the IMF of \citet{kroupa2001} and uniform pairing (Section~\ref{sec:org96ac7dd}).}
\end{figure}

\begin{figure}{}
\centering
\includegraphics[width=8.cm]{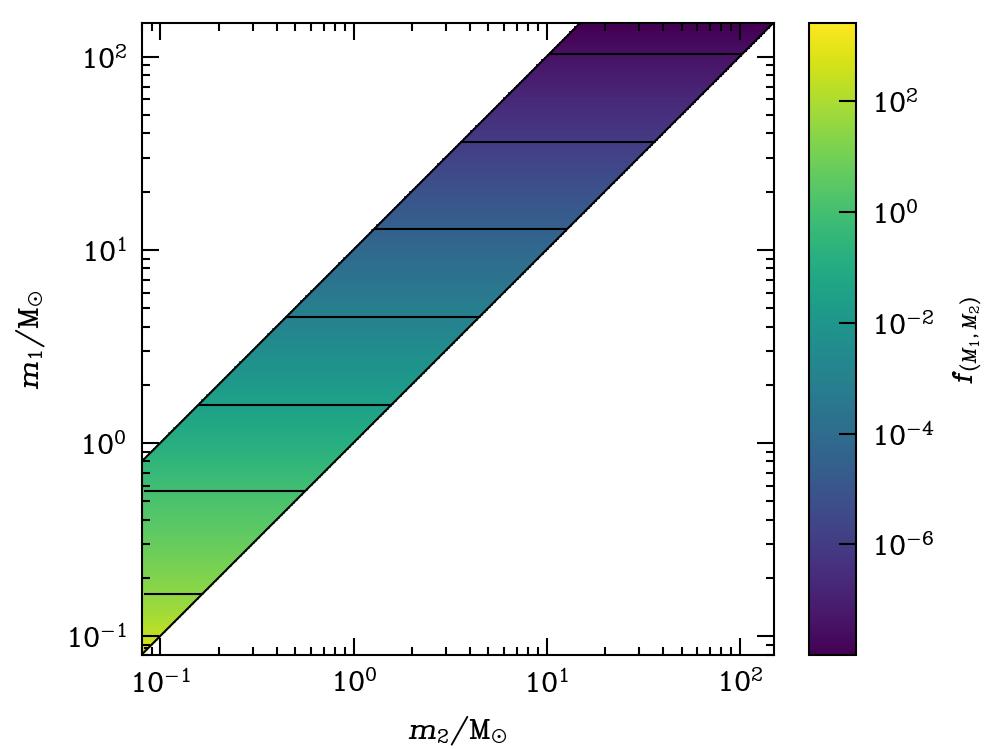}
\caption{\label{fig:orgf68edd4}The joint mass function, \(f_{(M_{\primary}, M_{\secondary})}\), constructed using the IMF of \citet{kroupa2001} and uniform pairing (Section~\ref{sec:org96ac7dd}, Figure~\ref{fig:org1e8a02d}). It is non-zero for \(m_{\secondary} \le m_{\primary}\) and \(m_{\secondary} \ge q_{\min}m_{\primary}\) where \(q_{\min} = 0.1\). As is the case for random pairing (Figure~\ref{fig:org0ab2b78}), low-mass pairs are most probable and high-mass pairs least probable.}
\end{figure}

Let us again suppose that \(f_{M} = \xi\) and, furthermore, that \(q_{\min} = 0.1\). We plot the PMF and SMF alongside the MF in Figure~\ref{fig:org1e8a02d} and the JMF in Figure~\ref{fig:orgf68edd4}. Note that the PMF has a maximum at \(m = \qty{0.11}{\solarmass}\) while the SMF and MF have maxima at \(m = \qty{0.08}{\solarmass}\). The three functions are equal at \(m = \qty{0.22}{\solarmass}\).

\section{Discussion}
\label{sec:org39c4229}

We have used the canonical IMF in combination with the CSMFs for random and uniform pairing to illustrate the application of our method. But we could have used any MF and CSMF (notwithstanding the fact that Equation~\ref{eq:fredholm_equation_for_primary_mass_function} must admit a non-negative solution). Neither random nor uniform pairing represents a physical population but our method allows the use of pairing schemes that do. In choosing a MF and CSMF for a particular binary-star population we wish to represent accurately the population of free stars from which the binary stars have been formed as well as the mechanism that has formed them. But even in the case of an initial population of binary stars these are uncertain. Theory does not yet provide a first-principles derivation of the IMF or initial CSMF meaning that phenomenological models of them are a necessity. Typically we pick from a library of formulas. For the MF we might pick one of the formulas suggested by \citet{kroupa2001}, \citet{chabrier2003}, or \citet{salpeter1955}. For the CMRF (from which we can derive the CSMFs) we might pick one of various piecewise powerlaws in \(q\) \citep{duchene2013,offner2023}.

These alternative MFs and CMRFs differ not only in their formulas but also in their support. As we have noted, the infimum and supremum of the support of the MF are \(m_{\min}\) and \(m_{\max}\). Similarly the infimum of the support of the CSMF is never less than \(m_{\min}/m_{\primary}\) and may be restricted to be also never less than \(q_{\min}\).
Together these values define the range of masses that are available for pairing. They are therefore crucial to our method. Changing the support of either the MF or the CSMF changes the resulting JMF. It is therefore worth discussing the possibility and potential consequences of using other, more
plausible, MFs and CSMFs.

\subsection{Mass function}
\label{sec:org7b2bbd8}

In specifying the MF (Equation~\ref{eq:kroupa_imf}) we have assumed that stars can be paired only with other stars and have therefore used the minimum possible stellar mass, namely the hydrogen burning limit, for \(m_{\min}\), which we have taken to be \(\qty{0.08}{\solarmass}\) \citep{whitworth2018}. But if stars can also be paired with brown dwarfs then \(m_{\min}\) would be the minimum possible brown-dwarf mass, namely the opacity limit. This depends weakly on temperature and strongly on mean molecular mass and is approximately \(\qty{0.003}{\solarmass}\) in the Solar neighbourhood \citep{whitworth2018}. Similarly, we have assumed there to be a maximum stellar mass, which we have taken to be \(\qty{150}{\solarmass}\) \citep{weidner2004,figer2005,oey2005}, although the mass of the most massive star is disputed and could be significantly greater, say \(\qty{300}{\solarmass}\) \citep{crowther2010}.

Our results for random and uniform pairing are sensitive to the choice of \(m_{\min}\) but insensitive to the choice of \(m_{\max}\). In both cases the gross features of the PMF and hence JMF are unchanged. The PMF vanishes in the low-mass limit and tends to twice the value of the IMF in the high-mass limit, as it must (Section~\ref{sec:orgce03346}). In both cases the PMF also retains its single mode. If \(m_{\min} = \qty{0.003}{\solarmass}\) the mode for random pairing reduces from \(\qty{0.16}{\solarmass}\) to \(\qty{0.0060}{\solarmass}\) whilst the mode for uniform pairing reduces from \(\qty{0.11}{\solarmass}\) to \(\qty{0.0041}{\solarmass}\). If \(m_{\max} = \qty{300}{\solarmass}\) the modes are effectively unchanged in both cases. The sensitivity of the PMF to \(m_{\min}\) and insensitivity to \(m_{\max}\) reflect that facts that, if the canonical IMF holds, low stellar masses are probable and high stellar masses improbable. Both theory and observation suggest that nuclear-burning stars are largely paired with other nuclear-burning stars \citep{goodwin2007,bate2015,offner2023,kroupa2024} so we must either set \(m_{\min}\) equal to the hydrogen burning limit or set \(m_{\min}\) to the opacity limit and use a CSMF that suitably disfavours the pairing of stars and brown dwarfs.

Just as important as the infimum of the MF's support is the MF's formula.
We have assumed that the formula of the MF is that suggested by \citet{kroupa2001} for the canonical IMF. The IMF of \citet{chabrier2003} would have done just as well although the IMF of \citet{salpeter1955} would not, since it is limited to high-mass (\(m > \qty{1}{\solarmass}\)) stars. The canonical IMF is the PDF of stellar mass at birth in the Solar neighbourhood. It describes the distribution of the masses of stars contained in stellar systems of all orders (unary, binary, ternary, etc.). But the distribution of masses in the population as a whole may differ from that in any subpopulation containing systems of one order. The IMF originates in the stars' formation mechanism \citep{hennebelle2024} and the formation of binary stars may result in an IMF that differs from the canonical IMF \citep{thies2007}. Binary systems do indeed appear to contain a deficit of low-mass stars \citep{offner2023} and the binary-star IMF may itself have a single mode at some intermediate mass. In these circumstances the modes of the PMFs for random and uniform pairing would both increase.

\subsection{Conditional secondary-mass function}
\label{sec:org34a4e35}

Both random and uniform pairing are popular on account of their simplicity.
For example, they are implemented by many population-synthesis software packages, including STARTRACK, MOCCA, BPAS, COMPAS, and binary\textunderscore{}c \citep{belczynski2008,hypki2013,eldridge2017,riley2022,izzard2023}.
Of course, to date, uniform pairing has only been used to construct approximations to the JMF, typically by means of primary-constrained pairing according to which the PMF is assumed to be equal to the IMF \citep{kouwenhoven2009}.

The true pairing scheme is significantly more complicated but remains elusive.
More sophisticated formulas for the initial CMRF are typically power laws or split-power laws \citep{duchene2013,offner2023}.
These are invariably restricted to small primary-mass ranges, typically to the mass ranges associated with one or two spectral classes.
Moreover, they are are subject to large uncertainties, especially in the case of binary systems with small primary-star masses or small mass ratios, observations of which are scarce.
\citet{offner2023} have given an excellent account of the state of the literature.

The problem of coverage even persists in the synoptic study of \citet{moe2017}, who compiled observations of zero-age main-sequence binary systems with primary stars of spectral class O to G with masses between \(\qty{0.8}{\solarmass}\) and \(\qty{40}{\solarmass}\) and mass ratios greater than \(0.1\). They did not give a CMRF explicitly but rather a conditional joint distribution of mass ratio, period, and eccentricity given primary mass, from which we may derive it. 
But this CMRF too is defined on these restricted ranges of primary mass and mass ratio and is therefore insufficient for our needs.

\section{Conclusion}
\label{sec:org6fe2d36}

We have seen how to find the JMF for a given MF and CSMF by first solving an integral equation to find the PMF. The JMF is then given by the product of the probability densities given by the PMF and CSMF. We have also considered the two special cases of random and uniform pairing assuming that the MF is identical to the IMF proposed by \citet{kroupa2001}. In both cases the PMF has a local maximum. The mass at which this maximum is located is smaller for uniform pairing than it is for random pairing. Accordingly, in uniform pairing there are more low-mass primary stars and fewer high-mass primary stars than there are in random pairing. Similarly, there are more high-mass secondary stars and fewer low-mass secondary stars.

Neither random nor uniform pairing are suitable for representing a physical population of binary stars but more plausible formulas for the CSMF are not defined for the full range of primary masses and mass ratios in the way that our method requires. Similarly, existing formulas for the MF fail to account for the deficit of low-mass stars in binary systems. We therefore need better models of the binary-star MF and CMRF to fully exploit our method. With these in place we might then revisit the problems of IMF inference, binary-star population synthesis, and their consequences for galaxy evolution.

\section*{Acknowledgements}
\label{sec:org94891c5}

AG and PD are supported by a UK Research and Innovation Future Leaders Fellowship (grant MR/S032223/1). RGI is funded by Science and Technology Facilities Council (grants ST/Y002350/1, as part of the BRIDGCE UK network, and ST/X005844/1).

\bibliographystyle{aasjournal}
\bibliography{bibliography}

\begin{appendix}
\refstepcounter{section} % Secretly moves the counter to 'A'
\label{sec:org5b48980}
  
Here we solve the Fredholm integral equation for the PMF (Equation~\ref{eq:fredholm_equation_for_primary_mass_function}) using uniform pairing, the CSMF of which is given by Equation~\ref{eq:uniform_csmf}. In this case,
\begin{align}
\begin{split}
&f_{M_{\primary}}(m)
= 2f_{M}(m) - \int_{m_{\min}}^{m_{\max}}\dfrac{\mathbf{1}_{B}(m_{\primary})f_{M_{\primary}}(m_{\primary})}{m_{\primary}(1 - \max(q_{\min}, m_{\min}/m_{\primary}))}\diff{}m_{\primary}
\end{split}
\end{align}
where \(\mathbf{1}_{B}\) is the indicator function of the interval \(B := [m, \min(m/q_{\min}, m_{\max})]\). This equation does not have closed-form solution but we can instead use a numerical method, specifically the Nyström method \citep{hackbusch1995}.

When using the Nyström method we approximate the integral using a quadrature rule with nodes \((m_{i})_{i = 0}^{n}\) and weights \((w_{i})_{i = 1}^{n}\). In this case we use product integration \citep{hackbusch1995}. We then approximate \(f_{M_{\primary}}\) by \(\hat{f}_{M_{\primary}}\) and write
\begin{align}
\hat{f}_{M_{\primary}}(m)
=
2f_{M}(m)
- \sum_{i = 0}^{n}w_{i}f_{M_{\secondary}|M_{\primary}}(m|m_{i})\hat{f}_{M_{\primary}}(m_{i}).
\end{align}
To find the elements of \((\hat{f}_{M_{\primary}}(m_{i}))_{i = 0}^{n}\) we let \(m = m_{j}\), whereupon
\begin{align}
\hat{f}_{M_{\primary}}(m_{j})
=
2f_{M}(m_{j})
- \sum_{i = 0}^{n}w_{i}f_{M_{\secondary}|M_{\primary}}(m_{j}|m_{i})\hat{f}_{M_{\primary}}(m_{i})
\end{align}
for all \(m_{j} \in (m_{i})_{i = 0}^{n}\), which we may write in matrix form as 
\begin{align}
\hat{\mathbf{f}}_{M_{\primary}} = 2\mathbf{f}_{M} - \mathbf{A}\hat{\mathbf{f}}_{M_{\primary}}
\end{align}
where \(\hat{\mathbf{f}}_{M_{\primary}} := (\hat{f}_{M_{\primary}}(m_{i}))_{i = 1}^{n}\), \(\mathbf{f}_{M} := (f_{M}(m_{i}))_{i = 1}^{n}\), and \(\mathbf{A} := (w_{i}f_{M_{\secondary}|M_{\primary}}(m_{i}, m_{j}))_{i, j = 1}^{n}\).
This is a system of linear equations,
\begin{align}
(\mathbf{I} + \mathbf{A})\hat{\mathbf{f}}_{M_{\primary}} = 2\mathbf{f}_{M},
\end{align}
with solution
\begin{align}
\label{eq:quadrature_solution_of_vie_ii}
\hat{\mathbf{f}}_{M_{\primary}} = 2(\mathbf{I} + \mathbf{A})^{-1}\mathbf{f}_{M}
\end{align}
if \(\mathbf{I} + \mathbf{A}\) is invertible, which in this case it is.
\end{appendix}
\end{document}